\documentclass{aa} 

\usepackage{caption}
\usepackage{graphicx}
\usepackage{txfonts}
\usepackage[colorlinks,citecolor=blue]{hyperref}
\usepackage{threeparttable}

\begin{document}

   \title{Investigating magnetically induced distortions of neutron stars through gamma-ray burst X-ray plateaus}

   \subtitle{}

   \author{Tingting Lin\inst{1,2}, Shuang Du\inst{3,1}, Weihua Wang\inst{3}, Shujin Hou\inst{4}, and Renxin Xu\inst{1,2} }

\institute{School of Physics, Peking University, Beijing 100871, China
         \and
              Kavli Institute for Astronomy and Astrophysics, Peking University, Beijing 100871, China
       \and
              College of Mathematics and Physics, Wenzhou University, Wenzhou Zhejiang 325035, China\\
              \email{dushuang@pku.edu.cn}
                     \and
                     Department of Physics and Electronic Engineering, Nanyang Normal University, Nanyang Henan 473061, China
             }



  \abstract
   {The magnetic field may distort neutron stars (NSs), but its effect has not yet been robustly tested through gravitational-wave observations due to the absence of a fast-rotating Galactic magnetar.
   The investigation of parts of gamma-ray bursts (GRBs) can potentially shed light on the magnetically induced distortion since their central objects may be millisecond magnetars.
   In this paper we propose a method for estimating the distortions of these possible magnetars under the GRB magnetar scenario.
   According to the case study of GRB 070521, we find a relation between the effective magnetically induced ellipticity, $\epsilon_{\rm B,eff}$, and the effective dipole magnetic field strength on NS surfaces, $B_{\rm eff}$, namely $\epsilon_{\rm B,eff}\sim 10^{-3}(B_{\rm eff}/10^{15}\rm G)^{2}$.
   Furthermore, we constrain the internal magnetic field structure of the magnetar to be $B_{\rm eff}\sim 0.02 <B_{\rm t}>$ and $B_{\rm eff}\sim 0.1B_{\rm t}$,
   where $<B_{\rm t}>$ is the volume-averaged internal toroidal field.
   This constraint can be used as the initial condition in modeling the structure of NS magnetospheres.
   Finally, the possibility of testing the method shown in this paper through gravitational-wave observations is discussed.}


\authorrunning{Lin et al.} \titlerunning{Investigating magnetically induced distortions of NSs through GRB X-ray plateaus}

   \maketitle

%

\section{Introduction}\label{sec1}
Rapidly rotating, distorted neutron stars (NSs) are potential gravitational-wave sources (e.g., \citealt{2014PhRvL.112q1102M,2016PASJ...68S..12M}).
Many authors have considered the influence of the anisotropic pressure of the internal magnetic fields in NSs,
 especially that of the internal toroidal magnetic field \citep{1969ApJ...157.1395O,1996A&A...312..675B,2000A&A...356..234K,2001MNRAS.327..639I,2001A&A...367..525P,2002PhRvD..66h4025C,2004ApJ...600..296I,2005ApJ...634L.165S,
2005MNRAS.359.1117T,2008MNRAS.385..531H,2009MNRAS.398.1869D,2011MNRAS.417.2288M}.
However, the gravitational radiation generated by distorted NSs has not been detected, even in the Milky Way \citep{2014ApJ...785..119A,2017PhRvD..96f2002A,2017ApJ...844..112G}.
The reason may be that Galactic NSs do not simultaneously maintain strong magnetic fields ($>10^{14}\;\rm G$) and fast spins (in LIGO's range),
which are needed to produce strong gravitational radiation.
Based on the observations of soft gamma repeaters and anomalous X-ray pulsars,
the surface magnetic dipole field strengths of Galactic magnetars are $\sim 10^{14}-10^{15}\;\rm G$ (see \citealt{2011ASSP...21..247R} for a review),
but spin periods of these magnetars are usually longer than $1\;\rm s$ due to the effective braking by magnetic dipole radiation.
Accretion may accelerate NSs, but it will in turn result in the decay of the magnetic field (e.g., \citealt{2016ApJ...833..261B,2016ApJ...833..189L,2019AN....340.1030G}),
so the magnetic field strengths of millisecond pulsars are usually weaker than those of normal pulsars (\citealt{2005AJ....129.1993M}).

To test the effect of magnetically induced distortion through the gravitational-wave channel, fast-rotating, strongly magnetized NSs (i.e., millisecond magnetars) are needed.
Using the dynamo theory, \cite{1992ApJ...392L...9D} showed that nascent NSs with millisecond periods can generate internal magnetic fields of up to $10^{16}\rm G$
as long as these NSs experience a phase of neutrino-driven turbulent convection\footnote{An alternative ``powerful contender'' is the fossil field scenario (\citealt{2006MNRAS.367.1323F,2008MNRAS.389L..66F}; see \citealt{2015RPPh...78k6901T} for a review).}.
So the key issue is what event can provide such a turbulent condition and reveal the existence of the millisecond magnetar.
We note that since gamma-ray bursts (GRBs) originate from massive star collapses \citep{1993ApJ...405..273W,1998ApJ...494L..45P, 1998Natur.395..670G, 1999ApJ...524..262M, 2003Natur.423..847H, 2003ApJ...591L..17S} or NS binary mergers \citep{1986ApJ...308L..43P,1987ApJ...314L...7G, 1989Natur.340..126E, 2017ApJ...848L..13A},
the catastrophes of GRB progenitors may provide the condition for forming millisecond magnetars,
that is to say, GRB prompt emissions and afterglows are potential events powered by nascent millisecond magnetars \citep{1992Natur.357..472U,1994MNRAS.267.1035U,1998PhRvL..81.4301D,1998A&A...333L..87D,KR,2001ApJ...552L..35Z,DWWZ,2010ApJ...715..477Y,2010MNRAS.409..531R,
2011MNRAS.413.2031M,2013MNRAS.430.1061R,2017MNRAS.468.3202B,2020ApJ...901...75D}.

It is currently difficult to directly detect gravitational radiation from a millisecond magnetar in a GRB \citep{2019ApJ...875..160A},
and thus an indirect way is needed.
Under the GRB magnetar scenario, some GRB X-ray afterglows (i.e., X-ray plateaus) are powered by magnetar winds, and
the time evolutions of the fluxes of these GRB X-ray afterglows are related to the spin-downs of these magnetars (the former must track the later; e.g., \citealt{2010ApJ...715..477Y,2010MNRAS.409..531R,2013MNRAS.430.1061R}).
During the coevolution, the braking torques exerted by gravitational radiation will affect magnetar spin-downs as well as the evolution of the X-ray afterglows.
Therefore, information regarding gravitational radiation may be extracted from these afterglows through their flux evolutions.
\cite{2016MNRAS.458.1660L} constrained the ellipticity of GRB magnetars under this framework,
but the degeneracy among relevant parameters is not eliminated and the ellipticity is constrained through an inequation.
Furthermore, \cite{2020ApJ...892L..34S} proposed that the ellipticity can be estimated according to the quasi-periodic oscillation in some X-ray afterglows
since distortions of magnetars will induce free precessions, but \emph{Swift's} X-Ray Telescope does not have sufficient time resolution.

In this paper we present an improved method for estimating the magnetically induced distortion under two assumptions:
(i) GRB X-ray plateaus correspond to energy releases of magnetar winds solely behind GRB jets \citep{2020ApJ...901...75D,2021ApJ...922..102H}
rather than being the result of magnetar winds being injected into GRB jets \citep{1998A&A...333L..87D,2001ApJ...552L..35Z}, and
(ii) the gravitational radiation of magnetars is mainly stimulated by magnetically induced distortions with constant ellipticities
(the evolution of the magnetic field in a short duration is neglected).
The merit of this method is that once the gravitational radiation from a GRB central magnetar is detected,
the results obtained from the gravitational-wave channel and the electromagnetic-wave channel can be double-checked.

The remaining part of this work is organized as follows.
We show our method for estimating the magnetically induced ellipticity under the GRB magnetar scenario in Sect. \ref{sec2}.
We present a case study of a \emph{Swift} GRB sample, GRB 070521, in Sect. \ref{sec3}.
The summary and discussion are given in Sect. \ref{sec4}.

\section{The method}\label{sec2}
Many factors can lead to the distortion of NSs and the further emission of gravitational waves.
For example, starquakes, accretion, and fast spinning can distort NSs since mountains or oscillations may be produced \citep{1999LRR.....2....2K,2021arXiv210907858G}.
In this paper we only consider the gravitational radiation generated by the magnetically induced distortion for the following reasons.
(1) Starquakes need time to accumulate stress in NS crusts,
but nascent NSs are very hot, such that the stars (even fluid bodies) have considerable plasticity in the $\sim 10^{3}\rm s$ after their births.
Therefore, the accumulation will be restrained.
(2) The magnetar winds from millisecond magnetars can put strong pressure on the possible fallback-accretion matter, and
hence the fallback accretion will be prevented \citep{2011ApJ...736..108P}.
(3) Gravitational waves emitted by oscillating NSs due to their fast spins are not detected, even in Milky Way \citep{2020ApJ...902L..21A,2021ApJ...913L..27A}.

The gravitational radiation produced by the magnetically induced distortion depends on the magnetic inclination and wobble angle (e.g., \citealt{2001PhRvD..63b4002C,M08,2011MNRAS.417.2288M}).
Since these angles cannot be determined, nor can the realistic magnetically induced distortion, we introduce an effective magnetically induced ellipticity, $\epsilon_{\rm B,eff}$,
such that the luminosity of gravitational radiation is (e.g., \citealt{PM,M08})
\begin{eqnarray}\label{1}
L_{\rm gw}(t)=\frac{32GI^{2}\epsilon_{\rm B, eff}^{2}\Omega(t)^{6}}{5c^{5}},
\end{eqnarray}
where the other parameters are the gravity constant, $G$, moment of inertia, $I$, speed of light, $c$, and angular velocity, $\Omega(t)$.
Similarly, we use an effective dipole magnetic field strength on the NS surface, $B_{\rm eff}$, to include the deviation from the realistic magnetic field. The
power of spin-down wind is then (e.g., \citealt{LL79,1983bhwd.book.....S})
\begin{eqnarray}\label{2}
L_{\rm em}(t)=\frac{B_{\rm eff}^{2}R^{6}\Omega(t)^{4}}{6c^{3}},
\end{eqnarray}
where $R$ is the equatorial radius.

\begin{figure}[h]
\centering
  \includegraphics[width=0.72\textwidth]{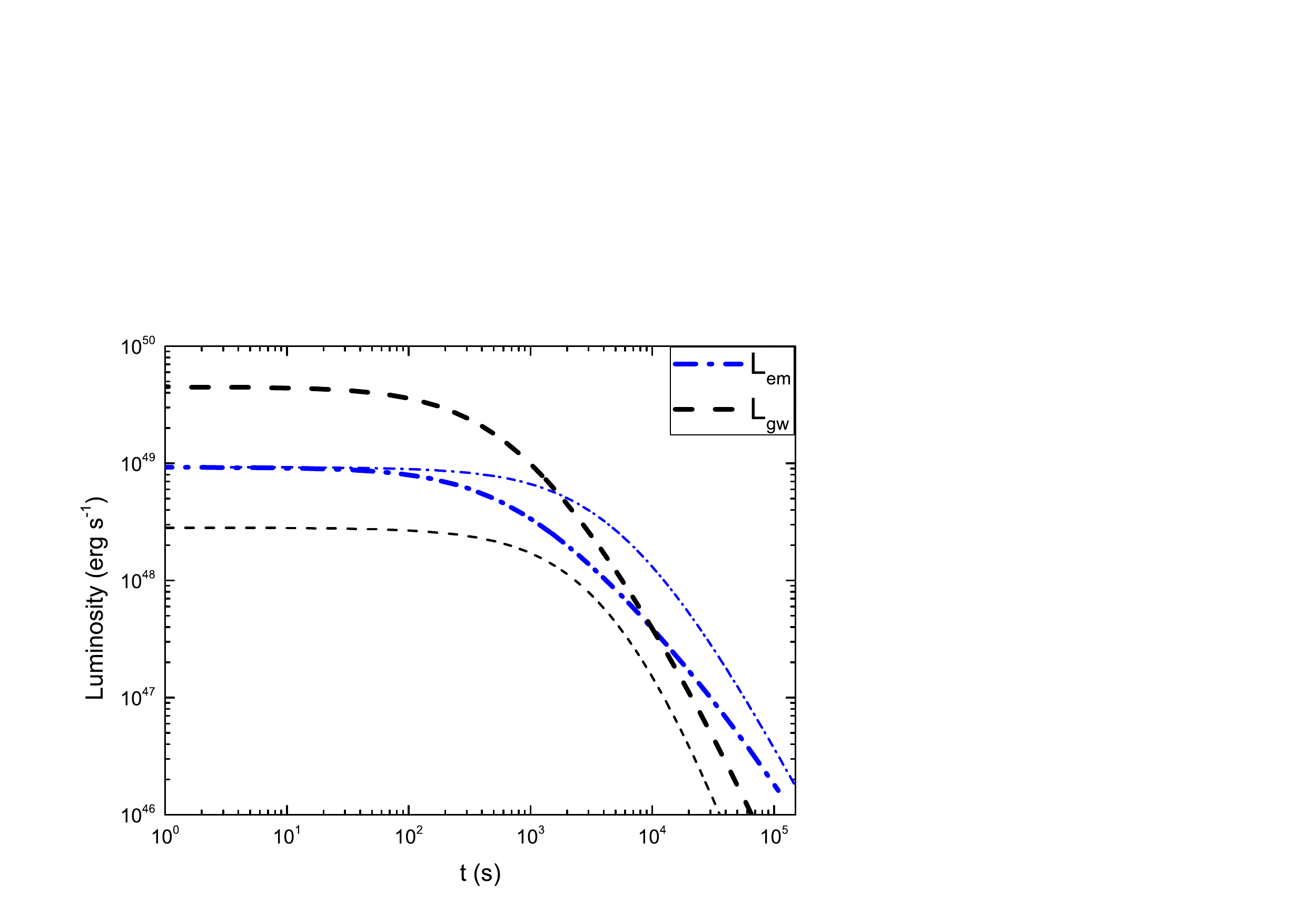}
  \caption{Time evolution of $L_{\rm em}$ and $L_{\rm gw}$.
           The thicker lines correspond to the parameters $\epsilon_{\rm B,eff}=2\times 10^{-3}$,
           $B_{\rm eff}=2\times 10^{14}\;\rm G$, $I=3\times 10^{45}\;\rm g\cdot cm^{2}$, and $\Omega_{0}=6280\;\rm rad\cdot s^{-1}$.
           The thinner lines correspond to the same parameters as those of the thicker lines, except $\epsilon_{\rm B,eff}=5\times 10^{-4}$.
           \label{fig2}}
\end{figure}

The general spin-down evolution can be obtained by solving the equation $I\Omega\dot{\Omega}=-L_{\rm em}-L_{\rm gw}$.
The solution is (\citealt{2016MNRAS.463..489H}; in which the $\gamma_{\rm e}$ in the logarithmic term is missing)
\begin{eqnarray}\label{eq17}
t=\left [ \frac{\gamma_{\rm e}}{2\beta ^{2}} \ln\left ( \frac{\beta +\gamma_{\rm e}\Omega(t)^{2}}{\gamma_{\rm e}\Omega(t)^{2}} \right )-\frac{1}{2\beta \Omega(t)^{2}}\right ]\bigg|_{\Omega }^{\Omega _{0}},
\end{eqnarray}
where
\begin{eqnarray}
\beta =\frac{B_{\rm eff}^{2}R^6}{6Ic^3},
\end{eqnarray}
\begin{eqnarray}
\gamma_{\rm e}=\frac{32GI\epsilon_{\rm B,eff}^{2}}{5c^5},
\end{eqnarray}
and $\Omega_{0}\equiv \Omega(t=0)$ (similarly hereinafter).
For comparison, two different situations -- in which the initial spin-down evolution is dominated by gravitational radiation (thicker lines)
and the initial evolution is dominated by magnetic dipole radiation (thinner lines) -- are shown in Fig. \ref{fig2}.
To see the difference between the two situations more clearly, it is convenient to consider two extreme cases \citep{1992Natur.357..472U,1998A&A...333L..87D,2001ApJ...552L..35Z,2016MNRAS.458.1660L}. First, when the gravitational radiation dominates the spin-down (i.e., $I\Omega\dot{\Omega}\approx -L_{\rm gw}$), according to Eqs. (\ref{1}) and (\ref{2}), the power of the spin-down wind is
\begin{eqnarray}\label{eq7}
L_{\rm em}(t)=L_{\rm em,0}\left ( 1+\frac{t}{\tau_{\rm gw}}{} \right )^{-1},
\end{eqnarray}
where
\begin{eqnarray}\label{8}
\tau_{\rm gw}=\frac{5c^{5}}{128GI\epsilon_{\rm B,eff}^{2}\Omega_{0}^{4}}.
\end{eqnarray}
Second, when the magnetic dipole radiation dominates the spin-down (i.e., $I\Omega\dot{\Omega}\approx -L_{\rm em}$),  according to Eq. (\ref{2}), the power of the spin-down wind is
\begin{eqnarray}\label{eq5}
L_{\rm em}(t)=L_{\rm em,0}\left ( 1+\frac{t}{\tau_{\rm em}}{} \right )^{-2},
\end{eqnarray}
where
\begin{eqnarray}\label{6}
\tau_{\rm em}=\frac{3c^{3}I}{B_{\rm eff}^{2}R^{6}\Omega_{0}^{2}}.
\end{eqnarray}

If the evolution of the GRB X-ray flux tracks that of the magnetar spin-down,
that is, the decay index of the segment following the GRB X-ray plateau is the same as that of the spin-down luminosity,
one can easily determine which energy-loss mechanism dominates the spin-down for the given extreme GRB sample according to the decay index: $\sim -1$ corresponds to gravitational-radiation domination (see Eq. \ref{eq7})
and $\sim -2$ corresponds to magnetic-dipole-radiation domination (see Eq. \ref{eq5}).
So, the distortion can be constrained through Eqs. (\ref{1}) and (\ref{2}).
For example, if the brake is initially dominated by gravitational radiation, there is (see also \citealt{2016MNRAS.458.1660L})
\begin{eqnarray}\label{3}
\epsilon_{\rm B,eff}>\left ( \frac{5B_{\rm eff}^{2}R^{6}c^{2}}{192GI^{2}\Omega_{0}^{2}} \right )^{1/2}
.\end{eqnarray}
Indeed, the coevolution between the GRB X-ray flux and the magnetar spin-down power is the basic requirement
as long as the "internal plateaus" \citep{TCO,2010MNRAS.402..705L,2010MNRAS.409..531R} can be explained under the GRB magnetar scenario \citep{2020ApJ...901...75D}.
However, Eq. (\ref{3}) is not sufficient to constrain $\epsilon_{\rm B,eff}$
since the parameters on the right side of this equation, $B_{\rm eff}$, $R$, $\Omega_{0}$, and $I$,  are uncertain.
The unknown parameters should be replaced by observable quantities.

We note that, under the magnetic-dipole-radiation-dominated case, two conditions should be satisfied \citep{2017arXiv171205964D}. (a)
The luminosity of the GRB X-ray afterglow powered by the spin-down wind, $L_{\rm x}(t)$, should be smaller than the spin-down power itself, that is,
\begin{eqnarray}\label{eq9}
L_{\rm x}(t)=\eta L_{\rm em}(t),
\end{eqnarray}
where $\eta \in (0, 1)$ is the efficiency with which the magnetic energy is converted into X-ray emission.
(b) The total energy of the GRB X-ray afterglow powered by the spin-down wind should be smaller than the rotational energy of the central magnetar, that is,
\begin{eqnarray}\label{eq10}
\int_{0}^{\tau_{\rm em}} L_{\rm x}(t)dt= \frac{1}{2}\eta I\Omega_{0}^{2}.
\end{eqnarray}
We note that $\eta$ is a constant since $L_{\rm x}(t)$ is approximatively a constant during the plateau segment (as is $L_{\rm em}(t)$ in $t\ll\tau_{\rm em}$);
this is just what the coevolution demands.
Combining Eqs. (\ref{eq5}), (\ref{eq9}), and (\ref{eq10}), one has
\begin{eqnarray}\label{eq11}
\Omega_{0}=\left (\frac{2\bar{L}_{\rm x}\tau_{\rm em}}{\eta I} \right)^{1/2},
\end{eqnarray}
where $\bar{L}_{\rm x}=\frac{1}{2}\eta L_{\rm em,0}$ is the average luminosity of the X-ray plateau.
Combining Eqs. (\ref{6}) and (\ref{eq11}), one has
\begin{eqnarray}\label{eq12}
B_{\rm eff}=\left (\frac{3\eta c^{3}I^{2}}{2\bar{L}_{\rm x}R^{6}\tau_{\rm em}^{2}}  \right )^{1/2}.
\end{eqnarray}

On the other hand, the power of gravitational radiation decays faster than that of the magnetic dipole radiation (see Eqs. (\ref{1}) and (\ref{2})).
When the spin-down is initially dominated by gravitational radiation,
there will be a moment, $\tau_{\ast}$, when the spin-down dominated by gravitational radiation
will transform into the case of magnetic-dipole-radiation domination \citep{2001ApJ...552L..35Z,2016MNRAS.458.1660L}.\ This gives
\begin{eqnarray}\label{eq13}
\tau_{\ast}=\frac{\tau_{\rm em}}{\tau_{\rm gw}}\left ( \tau_{\rm em}-2\tau_{\rm gw} \right ).
\end{eqnarray}
From Eq. (\ref{eq13}), $\tau_{\rm em}$ can be read as
\begin{eqnarray}\label{eq14}
\tau_{\rm em}=\tau_{\rm gw}+\sqrt{\tau_{\rm gw}^{2}+\tau_{\ast}\tau_{\rm gw}}.
\end{eqnarray}
The extreme sample of this case is that the decay index of the X-ray light curve changes as $0\rightarrow \sim -1\rightarrow \sim -2$ (see Eqs. (\ref{eq7}) and (\ref{eq5})).
If there is such a sample, one can obtain $\tau_{\rm gw}$ and $\tau_{\ast}$ through data fitting and get $\tau_{\rm em}$ through Eq. (\ref{eq14}). Then
$\epsilon_{\rm B,eff}$ can be directly connected to observable quantities via Eqs. (\ref{eq11}) and (\ref{eq12}).
Dividing Eq. (\ref{6}) by Eq. (\ref{8}), we get
\begin{eqnarray}\label{eq15}
\epsilon_{\rm B,eff}=\left (\frac{5\tau_{\rm em}B_{\rm eff}^{2}R^{6}c^{2}}{384\tau_{\rm gw}GI^{2}\Omega_{0}^{2}}  \right )^{1/2}.
\end{eqnarray}
Substituting Eqs. (\ref{eq11}) and (\ref{eq12}) into Eq. (\ref{eq15}), we get\begin{eqnarray}\label{eq16}
\epsilon_{\rm B,eff}=\sqrt{\frac{5Ic}{2G\tau_{\rm gw}}}\frac{\eta c^{2}}{16\tau_{\rm em}\bar{L}_{\rm x}}.
\end{eqnarray}
Therefore, the uncertain parameters on the right side of Eq. (\ref{3}) can be connected to observable quantities, except $\eta$ and $I$.

Previous work \citep{2017arXiv171205964D} has shown that $\eta\in \sim(0.01,1)$, but the efficiency still cannot be determined.
Here, we emphasize again that if the X-ray plateau is powered by the magnetar wind,
there must be a coevolution between the X-ray flux and the spin-down luminosity.
This demands that magnetic energy in the spin-down wind be almost totally released in X-ray emission and
 the energy of the magnetar wind solely be dissipated and not injected into the GRB jet, regardless of the dissipation mechanism (e.g., \citealt{1994MNRAS.267.1035U,2002A&A...391.1141D,2010ApJ...725L.234L,2011ApJ...726...90Z,2017MNRAS.468.3202B,2019ApJ...882..184L}).
This is because if the spin-down energy is injected into the jet, its dissipation will be modulated by the jet and the coevolution will be broken \citep{2020ApJ...901...75D}.
Therefore, we adopt the scenario in which GRB X-ray plateau emissions originate from the direct energy dissipation in the Poynting-flux-dominated magnetar winds
rather than dissipation through injecting magnetar winds into GRB jets (assumption (i) shown in the introduction).
According to assumption (i), the magnetic energy in the magnetar wind mainly transforms into X-ray emission to power the X-ray plateau; the remaining part changes into the kinetic energy of the wind and further transforms into X-ray emission to power an X-ray bump
through the interaction between the accelerated wind and the GRB jet (see Fig. 1 in \citealt{2020ApJ...901...75D}).
GRB 070110 \citep{TCO} fits this model exactly.
For GRB 070110, $\eta$ can be estimated as
\begin{eqnarray}
\eta=\frac{E_{\rm plateau}}{E_{\rm bump}+E_{\rm plateau}}\approx 0.9,
\end{eqnarray}
where $E_{\rm bump}$ is the total energy of the X-ray bump and $E_{\rm plateau}$ is the total energy of the X-ray plateau.

For simplicity, we adopted $\eta=0.9$ for all the chosen GRB samples in this paper since the efficiency mainly depends on the energy dissipation mechanism
(this is also the requirement of the GRB magnetar scenario, according to which most energy in magnetar winds should be released to power X-ray plateaus).
Once $I$ is known, $\epsilon_{\rm B,eff}$ can be estimated through Eqs. (\ref{eq14}) and (\ref{eq16}).
At present, the equation of state of NSs is still uncertain, as are the $R$ and $I$ of an NS with known mass.
However, observations have provided substantial information that allows us to estimate appropriate values for these two parameters
such that the result given by Eqs. (\ref{eq14}) and (\ref{eq16}) maintains rationality.
Observations confirm that GRB/type-Ic supernova associations \citep{1998Natur.395..670G, 2003Natur.423..847H, 2003ApJ...591L..17S}
indicate that long GRBs (duration $> 2\rm s$) originate from massive star collapses.
According to the mass distribution of Galactic pulsars (see \citealt{2016ARA&A..54..401O} for a review),
the mass of the millisecond magnetar associated with a long GRB could be $\sim 1.4\rm M_{\odot}$.
The direct measurement of NS masses \citep{2010Natur.467.1081D, 2013Sci...340..448A, 2019NatAs.tmp..472C}
and the observational and theoretical studies of binary NS mergers \citep{2017ApJ...848L..13A, 2017ApJ...850L..19M, 2018PhRvL.120q2703A, 2021arXiv210402070H}
show that the rotational inertia and equatorial radius of a millisecond NS with mass $\sim 1.4\;\rm M_{\odot}$
could be $I\sim 3\times 10^{45}\;\rm g\cdot cm^{2}$ and $R\sim 15\;\rm km$ \citep{2015PhRvD..92b3007C, 2019PhRvD..99d3004R}, respectively.

\section{A case study }\label{sec3}

\begin{figure}
\centering
  \includegraphics[width=0.43\textwidth]{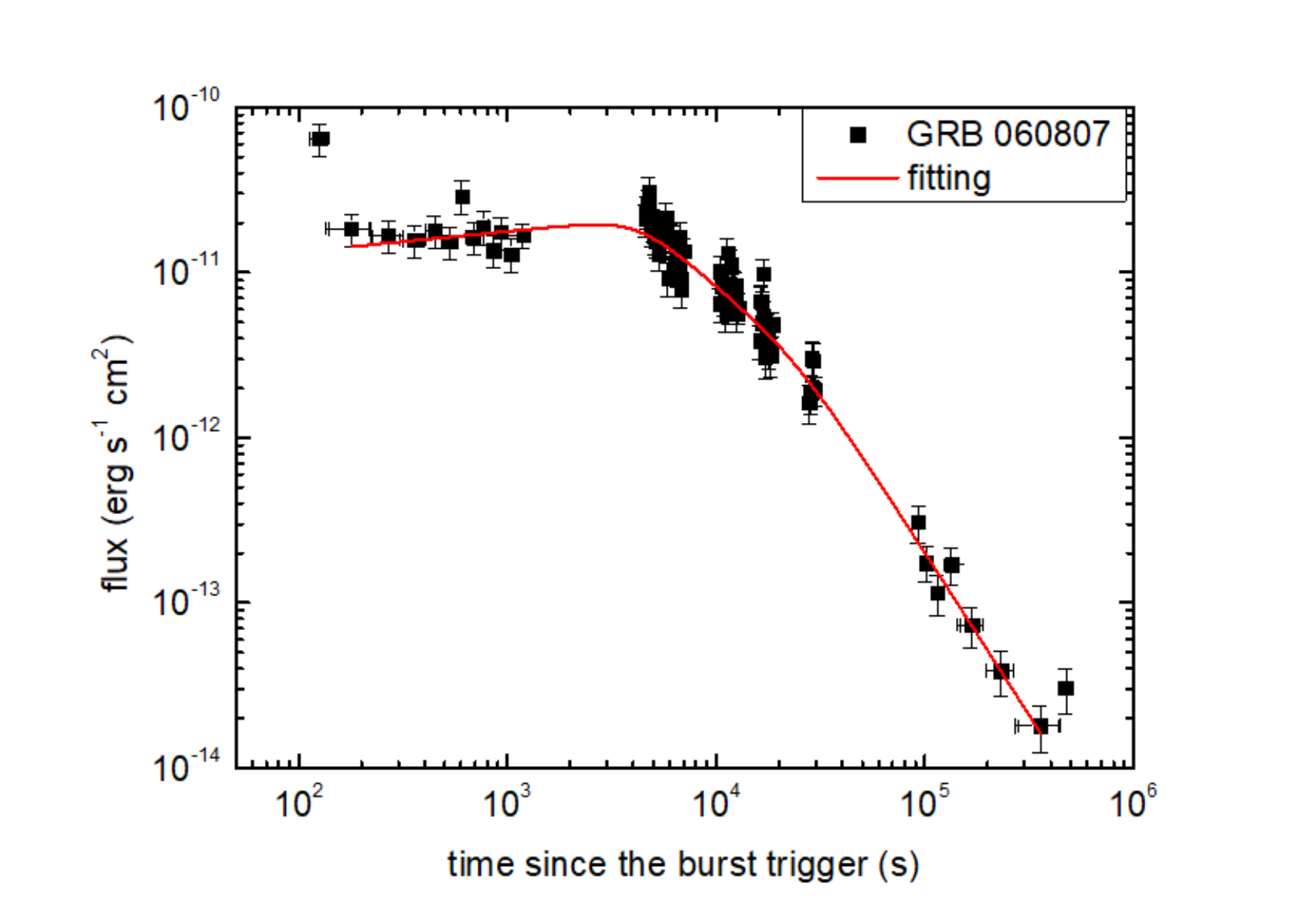}
  \includegraphics[width=0.43\textwidth]{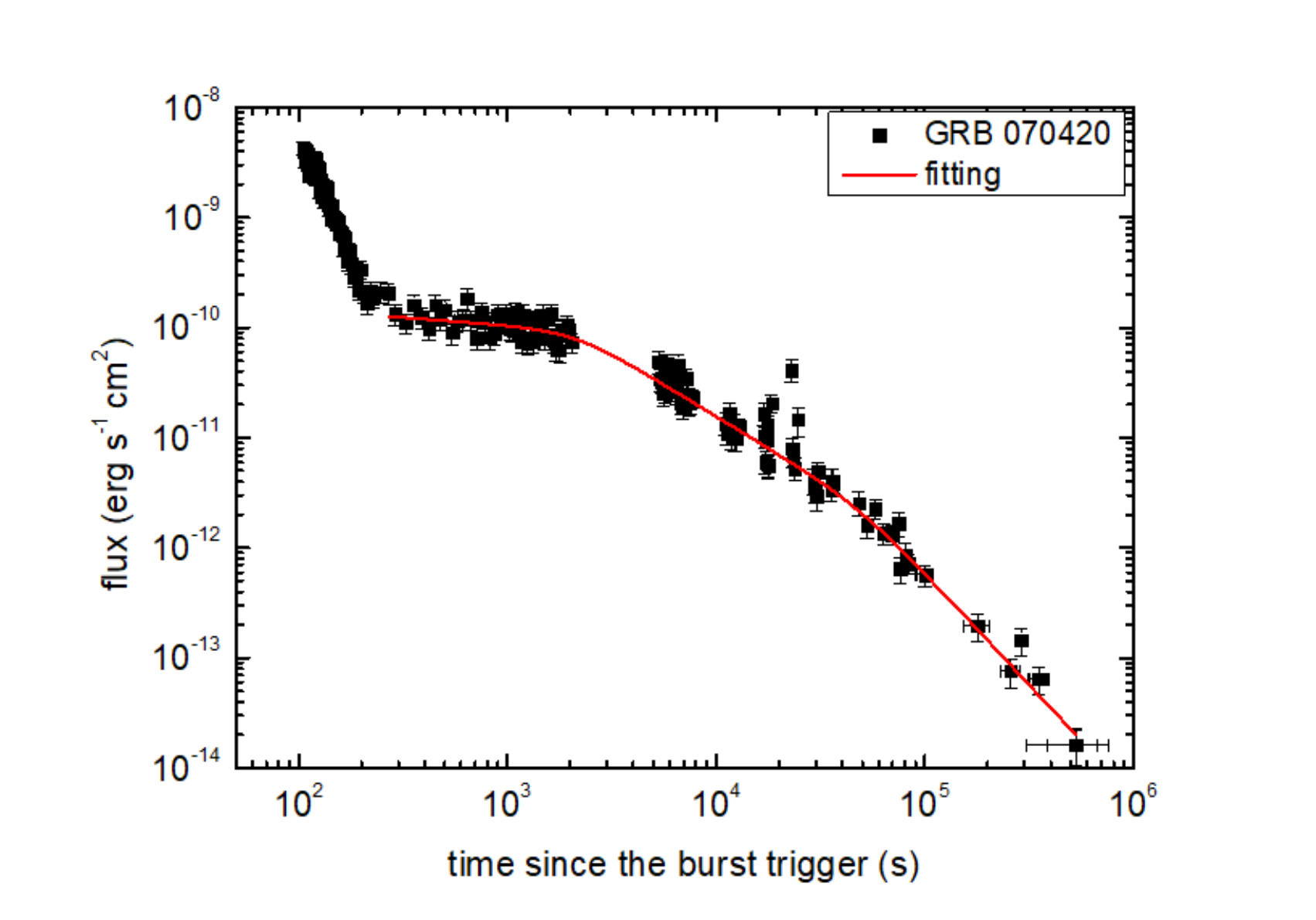}
  \includegraphics[width=0.43\textwidth]{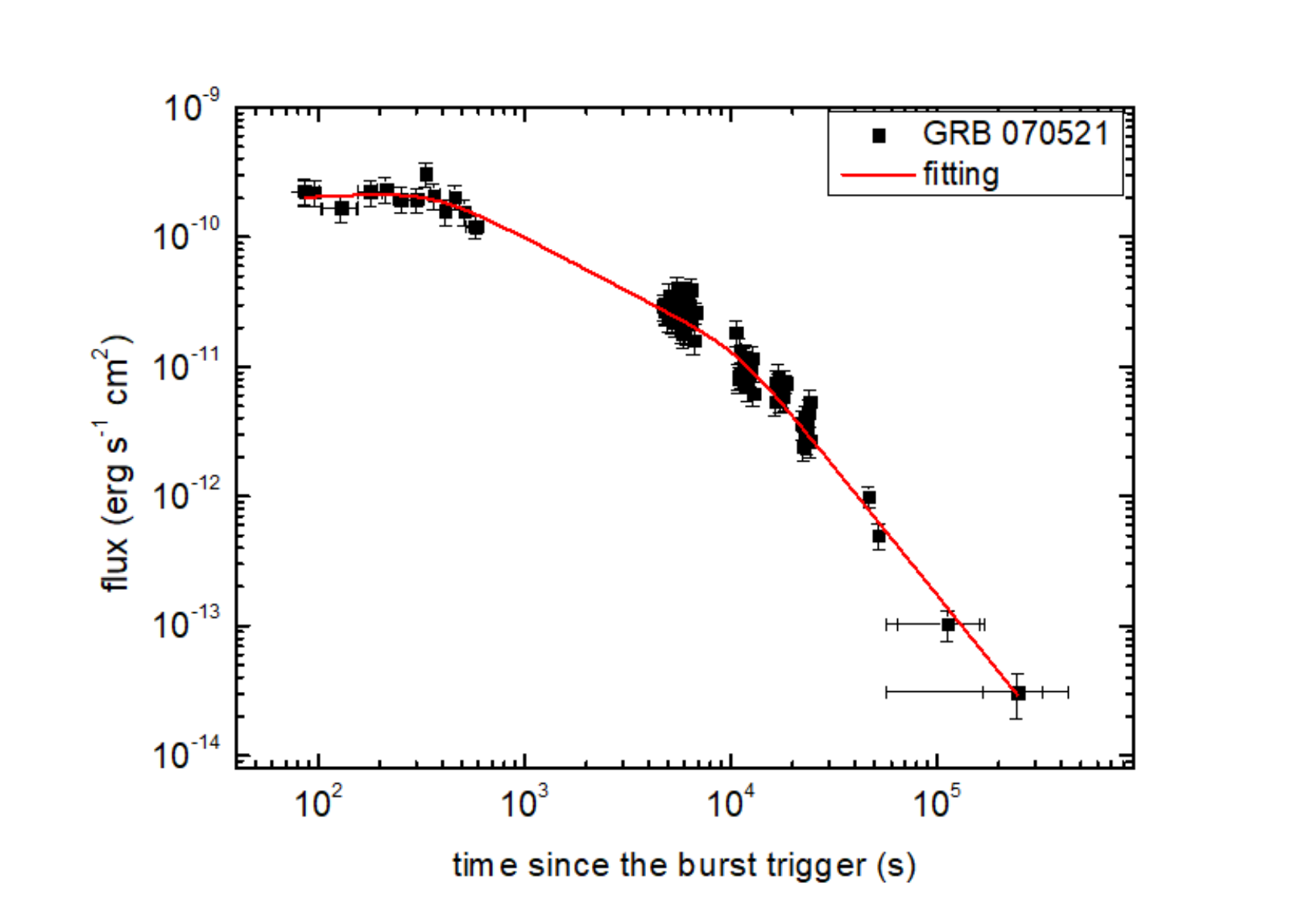}
  \caption{Fitting results of three GRB X-ray afterglows. \label{fig1}}
\end{figure}

\begin{table*}
\centering
\caption{Fitting result (the numbers in brackets are standard errors).}\label{ta1}
\begin{center}
\begin{tabular*}{15cm}{cccccccc}
\hline
Sample & $\omega$ & $F\; ({\rm erg s^{-1} cm^{2}})$ &  $\alpha_{1}$  &$\alpha_{2}$ & $\alpha_{3}$ & $\tau_{\rm gw} ({\rm s})$ & $\tau_{\ast} ({\rm s})$ \\
\hline
GRB 060807 & 4 & $2.1\times 10^{-11}$ & -0.12(0.03) & 1.07(0.12) & 1.96(0.14) & 4188(349) & 24251(9273) \\
GRB 070420 & 4 & $9.4\times 10^{-11}$ & 0.16(0.04) & 1.15(0.06) & 2.03 (0.14) & 2095(168) & 48271(11959)\\
GRB 070521 & 4 & $2.3\times 10^{-10}$ &  -0.09(0.11) & 0.83(0.06) & 1.98(0.09) & 364(67) & 11091(1399) \\
\hline
\end{tabular*}
\end{center}
\end{table*}

The GRB samples that match the expectation shown in Sect. \ref{sec2} are rare (especially with known redshift).
According to \emph{Swift} data \citep{2007A&A...469..379E,2009MNRAS.397.1177E}, we find three samples (see Fig. \ref{fig1}).
We adopted the following function to fit the data (e.g., \citealt{2010ApJ...715..477Y}):
\begin{eqnarray}
Flux=F\left[ \left(\frac{t}{\tau_{\rm gw}}\right)^{\omega\alpha_1}+\left(\frac{t}{\tau_{\rm gw}}\right)^{\omega\alpha_2}+\left(\frac{\tau_{*}}{\tau_{\rm gw}}\right)^{\omega\alpha_2}\left(\frac{t}{\tau_{*}}\right)^{\omega\alpha_3}\right]^{-1/\omega}
.\end{eqnarray}We note that $F$ is only the observed average flux of the plateau as long as the plateau is absolutely horizontal (i.e., $\alpha_{1} =0$).
During the fitting, we changed the smooth factor, $\omega$ (usually an integer is adopted), to let $F$ get close to the average flux of the plateau (i.e., $F$  and $t_{\rm gw}$ must match each other well). Therefore, all the parameters can be constrained.
The fitting result is shown in Table 1.

From the three samples,
GRB 070521 was selected (a long GRB with duration $T_{90}\approx 37.9\rm s$; \citealt{PBC}) to estimate $\epsilon_{\rm B,eff}$ since only the redshift of this GRB
is known ($z=2.087$; \citealt{2000A&A...356..234K}).
Through Table 1, we averaged the unabsorbed X-ray flux (the intergalactic extinction is considered) of GRB 070521 in $364\rm s$
and get $\bar{F}_{\rm lux}({\rm GRB\; 070521})\approx 3.1\times 10^{-10}\rm erg\; cm^{-2}\; s^{-1}$.
Under the $\Lambda$ cold dark matter model with parameters that the Hubble constant $ H_{\rm 0}\approx 70\rm km\; Mpc^{-1}\; s^{-1}$,
the energy fraction of matter $\Omega_{\rm m}\approx 0.3,$ and the energy fraction contributed by vacuum $\Omega_{\rm \Lambda}\approx 0.7$ \citep{2016A&A...594A..13P},
we get $\bar{L}_{\rm x}(\rm GRB\; 070521)\approx 4.3\times 10^{48}\rm erg\; s^{-1}$.
According to Eqs. (\ref{eq14}) and (\ref{eq16}) and Table 1,  $\epsilon_{\rm B,eff}({\rm GRB\;070521})\sim 4\times 10^{-3}$.
Furthermore, according to Eqs. (\ref{eq12}) and (\ref{eq14}) and Table 1, $B_{\rm eff}({\rm GRB\;070521})\sim 1\times 10^{15}\rm G$.

\cite{2002PhRvD..66h4025C} predicted that the ellipticity, $\epsilon_{\rm t}$,
induced by the internal toroidal field, $B_{\rm t}$, satisfies (the minus is hidden)
\begin{eqnarray}
\epsilon_{\rm t}=\left\{\begin{matrix}
1.6\times 10^{-6}\left ( \frac{<B_{\rm t} > }{10^{15}\rm G} \right ) & B_{\rm t}<B_{\rm cl}\\
1.6\times 10^{-6}\left ( \frac{<B_{\rm t}^{2} > }{10^{30}\rm G^{2}} \right ) & B_{\rm t}>B_{\rm cl}
\end{matrix}\right.
,\end{eqnarray}
where $<\cdots>$ means volume-averaged over the NS interior, and $B_{\rm cl}$ is the critical field.
Therefore, if the two conclusions are compatible, considering the stability of the magnetic field structure ($B_{\rm eff}/B_{\rm t}$ should not be too small), $\epsilon_{\rm B,eff}\sim 10^{-3}\left (\frac{B_{\rm eff}}{10^{15}\rm G}  \right )^{2}$ (i.e.,
$B_{\rm cl}<10^{15}\rm G$, $B_{\rm eff}\sim 0.02 <B_{\rm t}>$).
We note that $B_{\rm t}$ in the interior of the star is usually much larger than that of the exterior of the star;
there may be $B_{\rm eff}\sim 0.1B_{\rm t}$ near the magnetar outer crust.
This is consistent with the requirement in modeling anomalous X-ray pulsars/soft gamma repeaters (\citealt{1993ApJ...408..194T,1995MNRAS.275..255T})
that the toroidal fields can be up to $10^{16}\;\rm G$  in the outer crusts of magnetars.

To test the above result, gravitational-wave observation is essential. It is beneficial to estimate the detectability of this kind of gravitational-wave signal.
Since the rotational periods of GRB magnetars are usually close to the limit of Keplerian rotation, the frequency of the gravitational wave is $f_{\rm gw}\sim 2\rm kHz$.
Therefore, the amplitude can be estimated as (see, e.g., \citealt{M08})
\begin{eqnarray}
h&=&\frac{4\pi^{2}G}{c^{4}}\frac{If_{\rm gw}^{2}}{D_{\rm l}}\epsilon_{\rm B,eff} \nonumber\\
&\approx& 5\times 10^{-23}\left ( \frac{\epsilon_{\rm B,eff} }{4\times 10^{-3}} \right )\left ( \frac{I}{3\times 10^{45}\rm{g\;cm^{2}}} \right )\nonumber\\
&&\times \left ( \frac{1\rm{Mpc}}{D_{\rm l}} \right )\left ( \frac{f_{\rm gw}}{2\rm{kHz}} \right )^{2},
\end{eqnarray}
where $D_{\rm l}$ is the luminosity distance and $G$ is the gravity constant.
The sensitivity limits for advanced LIGO and the Einstein Telescope at $2\rm kHz$ are $\sim 1.3\times 10^{-23}$ and $\sim 1.0\times 10^{-24}$ (e.g., \citealt{2010CQGra..27s4002P}), respectively.
This kind of gravitational-wave signal can be detected in several megaparsecs.
We note that the signal-to-noise ratio of the periodic signal can be enhanced by folding the data around the period;
the gravitational wave is expected to be detected at a greater distance
(using the typical duration of the X-ray plateau, $t_{\rm gw}$, as the standard, the multiple factor is $\sqrt{\tau_{\rm gw}}\sim 30 $).

\section{Summary and discussion}\label{sec4}
{In this paper we aim to investigate how strongly an NS can be distorted by a magnetic field.
We propose a method for estimating magnetically induced distortions of NSs through GRB X-ray plateaus under the GRB magnetar scenario.
If the distortion is strong enough, the corresponding gravitational radiation can dominate the initial NS spin-down and
affect the time evolution of the NS electromagnetic radiation (i.e., the decay index of the X-ray flux may change as $\sim 0\rightarrow \sim 1\rightarrow \sim 2$).
Therefore, the ellipticity may be connected to some measurable and inferred parameters.
According to the case study of GRB 070521,
we find that $\epsilon_{\rm B,eff}\sim 10^{-4} (\frac{B_{\rm eff}}{10^{14}\rm G})^{2}$.
Comparing this with previous well-accepted theoretical results, for example, $\epsilon_{\rm B}\sim 10^{-4} (\frac{B_{\rm t}}{10^{16}\rm G})^{2}$ \citep{2002PhRvD..66h4025C},
our result shows that there should be $B_{\rm eff}\sim 0.02 <B_{\rm t}>$ and $B_{\rm eff}\sim 0.1B_{\rm t}$ near the magnetar outer crust if the two results are compatible.
This deduction could be used as the initial internal magnetic field structure to model the external magnetosphere of an NS.

A large toroidal field ($\sim 10^{16}\rm G$) may induce a certain instability for the given magnetic configuration \citep{Akgun},
but the appearance of such a strong toroidal field seems to be inevitable \citep{2013MNRAS.435L..43C}.
Due to the observation of Galactic magnetars \citep{2011ASSP...21..247R}
and the model-independent constraint on the magnetic field for the GRB magnetar scenario \citep{2017arXiv171205964D},
we tend to believe $B_{\rm eff}\leq\sim 10^{15}\rm G$.
This leads to a disparity between $B_{\rm eff}$ and $<B_{\rm t}>$ under the GRB magnetar scenario.
The reason may be that the asymmetrical collapses and explosions of long GRB progenitors (e.g., GRB 070521)
induce extra torques, which results in differential rotations between the interiors and exteriors of proto-NSs to amplify the internal toroidal fields.
In the future, several supermassive star collapses and rare binary NS mergers in $\sim 100\rm Mpc$ \citep{2017ApJ...851L..16A} may provide opportunities to test the results shown in this paper.


%

\begin{acknowledgements}
We thank the anonymous referee for very useful comments
that have allowed us to improve our paper.
 We acknowledge the use of the public data from the Swift data archives.
This work is supported by the National SKA Program of China No. 2020SKA0120100, 
the National Key R\&D program of China No. 2017YFA0402602 and the strategic Priority Research Program of CAS (XDB23010200)..
\end{acknowledgements}

\end{document}